\documentclass[9pt,conference]{IEEEtran}
\newcommand{\quotes}[1]{``#1''}
\usepackage{cite}
\usepackage{amsmath,amssymb,amsfonts}
\usepackage{algorithmic}
\usepackage{graphicx}
\usepackage{textcomp}
\usepackage{xcolor}
\usepackage{amsmath,graphicx,booktabs}
\usepackage{amssymb}
\usepackage{hyperref}
\usepackage{microtype}
\def\BibTeX{{\rm B\kern-.05em{\sc i\kern-.025em b}\kern-.08em
    T\kern-.1667em\lower.7ex\hbox{E}\kern-.125emX}}
\begin{document}

\title{A Non-autoregressive Model for Joint STT and TTS}
\author{\IEEEauthorblockN{Vishal Sunder\IEEEauthorrefmark{1}, Brian Kingsbury\IEEEauthorrefmark{2}, George Saon\IEEEauthorrefmark{2}, Samuel Thomas\IEEEauthorrefmark{2}, Slava Shechtman\IEEEauthorrefmark{3}}\IEEEauthorblockN{Hagai Aronowitz\IEEEauthorrefmark{3}, Eric Fosler-Lussier\IEEEauthorrefmark{1} and Luis Lastras\IEEEauthorrefmark{2}}\IEEEauthorblockA{\IEEEauthorrefmark{1}The Ohio State University, Columbus, OH, USA} \IEEEauthorblockA{\IEEEauthorrefmark{2}IBM Research, Yorktown Heights, NY, USA}\IEEEauthorblockA{\IEEEauthorrefmark{3}IBM Research, Israel}}


\maketitle

\begin{abstract}
In this paper, we take a step towards jointly modeling automatic speech recognition (STT) and speech synthesis (TTS) in a fully non-autoregressive way. We develop a novel multimodal framework capable of handling the speech and text modalities as input either individually or together. The proposed model can also be trained with unpaired speech or text data owing to its multimodal nature. We further propose an iterative refinement strategy to improve the STT and TTS performance of our model such that the partial hypothesis at the output can be fed back to the input of our model, thus iteratively improving both STT and TTS predictions. We show that our joint model can effectively perform both STT and TTS tasks, outperforming the STT-specific baseline in all tasks and performing competitively with the TTS-specific baseline across a wide range of evaluation metrics.

\end{abstract}


\section{Introduction}
In the recent past, there has been a lot of work in developing large multimodal models which can handle both speech and text modalities~\cite{ren2019almost, ao2022speecht5, chen2023lauragpt, barrault2023seamlessm4t, das2024speechverse, wang2024viola, speechteam2024funaudiollm}. The underlying philosophy of this line of research is the development of models that can perform a variety of speech-to-text and text-to-speech tasks like automatic speech recognition (ASR), speech synthesis (TTS), speech-to-speech translation (S2ST), speech-to-text translation (S2TT), and spoken language understanding (SLU). There are two ways of accomplishing this. First, we can take a modular approach where individual models are built for speech and text processing tasks using large amounts of data and then combined in a way that the above multimodal tasks are performed by utilizing a specialized combination of the modality-specific modules. SeamlessM4T~\cite{barrault2023seamlessm4t} and FunAudioLLM~\cite{speechteam2024funaudiollm} take this approach. The second approach aims at training a single model that can handle all tasks with weight sharing across modalities~\cite{ao2022speecht5, chen2023lauragpt, wang2024viola}.

The benefit of the first set of approaches is that models for individual tasks are easier to train, but the disadvantage is that combining them leads to an increase in model size and the tasks may not be aligned with each other. This disadvantage can be overcome using the second approach; yet, it can be challenging to train a model that can take both speech and text as input. Another feature of all the above models is that they are autoregressive; thus, their inference speed can be slow and they have a tendency to hallucinate~\cite{peng2024owsm}. Non-autoregressive (NAR) approaches for ASR~\cite{graves2006connectionist} and TTS~\cite{ren2019fastspeech} are significantly faster during inference and do not hallucinate on account of being frame synchronous. Indeed, large NAR models have been introduced in Peng et al.~\cite{peng2024owsm}, but they only perform speech-to-text tasks like ASR and S2TT. In this work, we explore whether we can perform speech recognition (STT) and speech synthesis (TTS) using a single multimodal model with a NAR approach. 


\noindent \textbf{Our contributions}: To the best of our knowledge, this is the first work that explores the possibility of performing STT and TTS jointly in a fully NAR manner. We utilize a CTC-based approach for STT and a FastSpeech-based approach for TTS. Our proposed multimodal model can take as input a sequence of text as a CTC alignment, a sequence of log-mel filterbank features, or a combination of both. The TTS task is done using a CTC alignment as input, which is another novelty of this work. We also train the model not only to perform STT and TTS, but also a set of tasks which use unpaired speech or text data for self-supervised learning and combined speech-text input for additional supervised learning tasks. Furthermore, we also propose an iterative refinement procedure during inference which effectively utilizes the multimodal nature of our model to refine STT and TTS predictions using a mask-predict approach.

\noindent \textbf{Relation with previous work}: As mentioned before, many models have been proposed in the literature which perform STT and TTS tasks jointly, but all of them do so autoregressively. The SpeechT5~\cite{ao2022speecht5} model comes closest to our proposed model in that it performs STT, TTS and self-supervised tasks. However, this model is autoregressive and after pretraining, needs to be adapted for individual tasks separately. We train our model jointly for all the tasks and do not require separate adaptation. Regarding the use of text-only unpaired data for STT improvement, two important techniques were proposed by Sainath et al.~\cite{sainath2023joist} and Thomas et al.~\cite{thomas2022integrating} which use token repetition and a so-called textogram, which is also repetition based, respectively, as the text modality input. Our model utilizes a masked CTC alignment for text-only data augmentation, which is a consistent input for the TTS task, and is completely non-autoregressive contrary to previous approaches. Semi-autoregressive STT has also been proposed recently to improve streaming STT~\cite{arora2024semi}, but our proposed iterative refinement approach is related more with recent work by Chi et al.~\cite{chi2020align} and Higuchi et al.,~\cite{higuchi2020mask, higuchi2022bert, higuchi2023bectra, higuchi2023mask}, for improving STT performance. However, our model is multimodal, i.e. it can take both speech and text as input, which allows the speech-text combined representation to be updated through iterations. In previous approaches, the speech representation from the encoder is fixed while the text from a decoder is updated and refined. Furthermore, our proposed iterative refinement can also work for TTS, which is completely novel to the best of our knowledge.

We show that it is possible to train a joint STT and TTS model in a NAR fashion. Although we find that the two tasks are not complementary to each other, using unpaired data during training and iterative refinement during inference leads either to parity with or better performance than unimodal baselines.

\section{Model design}
\label{sec:model_design}

\begin{figure}
    \hfill
    \centering
    \centerline{\includegraphics[width=0.8\columnwidth]{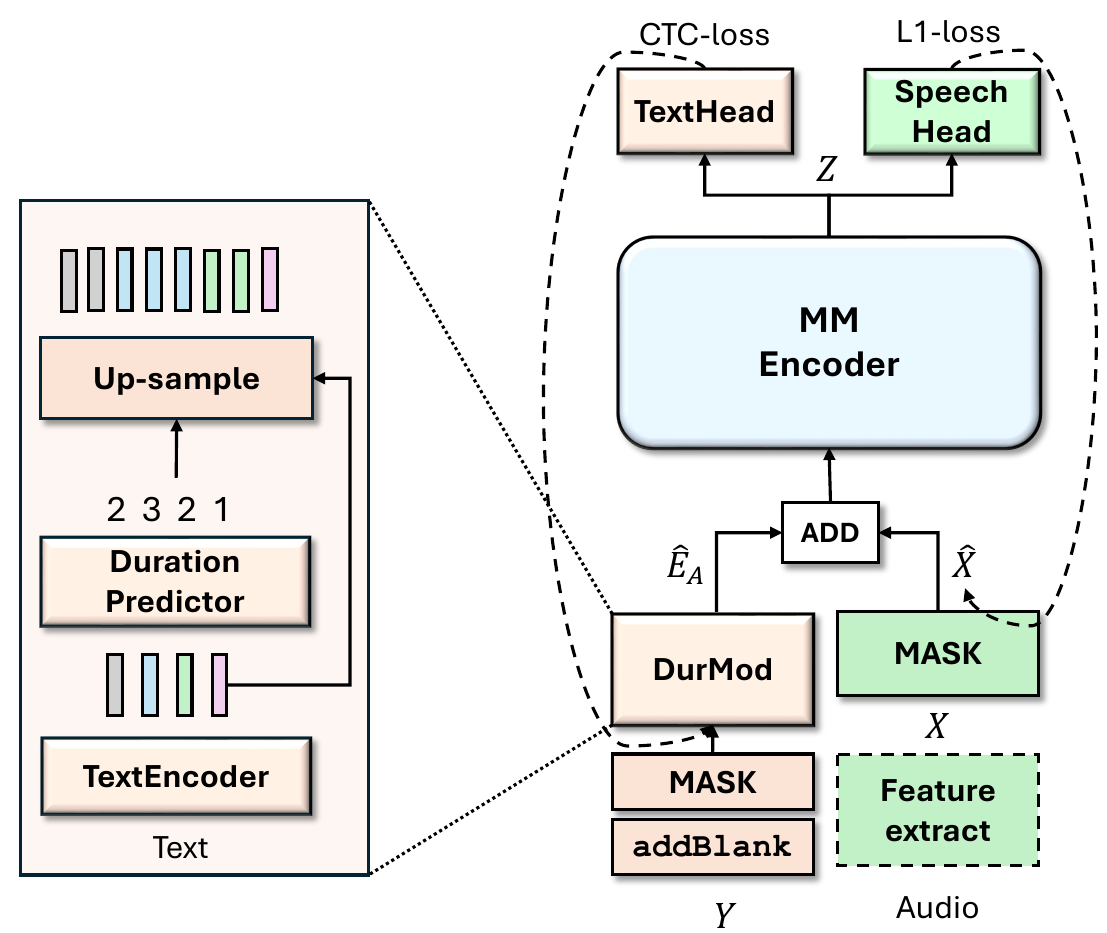}}
\caption{Model overview. The dashed arrows from the output back to the input represent the behavior during iterative refinement as explained in Section~\ref{subsec:iterative refinement}.} 
\label{fig:model}
\end{figure}

Our proposed model design is such that it can take speech, text, or a combination of both features as input. Let $X$ be a sequence of log-mel filterbank (LFB) features and $Y$ be the corresponding transcript. The multimodal model consists of four major components as shown in Figure~\ref{fig:model} and outlined below. 

\subsection{Components}
\label{subsec:components}
\noindent \textbf{Duration model}: The duration model converts the text $Y$ to a CTC alignment representation $E_A$. To do this, each character in $Y$ is interleaved with blanks (e.g. \quotes{CAT} is converted to \quotes{\_C\_A\_T\_}) and fed to a bidirectional text encoder to get character embeddings. These embeddings are repeated $R$ times\footnote{Here we use $R$ to represent a sequence of repetitions}, based on the ground truth alignment at train time. At test time, we utilize the repeat predictions ($\hat{R}$) from a duration predictor. Formally, this process is defined below,
\begin{align*}
    \begin{split}
        E_Y &= \text{TextEncoder}(\text{Embedding}(\texttt{addBlank}(Y))) \\
        E_A &= \text{Repeat}(E_Y, R) \\
        \hat{R} &= \text{DurationPredictor}(E_Y) \\ 
        L_{\text{dur}} &= \text{CrossEntropy}(\hat{R}, R)
    \end{split}
\end{align*}
Here, TextEncoder$(.)$ and DurationPredictor$(.)$ are both modeled as 2-layer conformers; however, DurationPredictor$(.)$ has an additional prediction layer on top. $L_{\text{dur}}$ is the loss used to train the model to predict the number of repetitions. Note that $E_A$ is just an upsampled version of $E_Y$. This part of the model is very similar to FastSpeech~\cite{ren2019fastspeech}, except that we use character-based CTC alignments instead of a phoneme alignment. The composite model (left side of Figure~\ref{fig:model}) is noted as DurMod(.) in the remainder of the paper.


\noindent \textbf{Masking}: We need a masking schedule for both speech and text modalities so that we can perform self-supervised learning with unpaired data and train the model for iterative refinement at inference.

To mask text, we randomly replace $p\%$ of the characters with a \texttt{<mask>} token. When passing this through the duration model, we replace a blank token with \texttt{<mask>} if the prior character is also masked. Thus, we get noisy upsampled CTC alignment embeddings, $\hat{E_A}$. We do not mask the first blank token. For example, if the CTC alignment of \quotes{CAT} is \quotes{\_CCA\_T\_} and its masked version is \quotes{C\texttt{<mask>}T}, $\hat{E_A}$ will correspond to \quotes{\_CC\texttt{<mask>}\texttt{<mask>}T\_}. This module is represented as $\text{Mask}_{Y}(.,p)$.

When masking speech, we use two types of schedules. First, we use a wav2vec2.0~\cite{baevski2020wav2vec} masking schedule where we randomly sample $p\%$ of log-mel frames as the starting frames without replacement and mask $M$ subsequent frames for each of the starting frames sampled. Masking a log-mel frame simply corresponds to replacing it with a zero-vector. This masking module is represented as $\text{Mask}_{X1}(.,p,M)$.

The second masking schedule for speech is useful for iterative refinement of TTS predictions. If the number of time steps is $T$ and the number of filters is $F$, we choose $p\%$ of $T$ and $F$ to get $t_0$ and $f_0$. Then, we mask all elements in $X$ along the time axis at $t = [t_0, T]$ and along the frequency axis at $f = [f_0, F]$. This masking module is represented as $\text{Mask}_{X2}(.,p)$.

\noindent \textbf{Multimodal encoder}: The multimodal encoder forms the main component of our model design and contains the majority of the parameters in the system. The multimodal encoder takes as input (possibly masked) text embeddings and (possibly masked) log mel features; this enables the model to variously handle speech-only, text-only, or combined input.  The two streams are added together as input to the model.
\begin{align*}
    \begin{split}
        \hat{X} &= \text{Mask}_{X1}(X,p_X,M) \\
        \hat{E_A} &= \text{DurMod}(\text{Mask}_{Y}(Y,p_Y)) \\
        Z &= \text{MMEncoder}(\hat{X} + \hat{E_A})
    \end{split}
\end{align*}
If $p_X = 1.0$, it implies that the speech modality is absent and $\hat{X}$ is a sequence of zero-vectors per the definition of $\text{Mask}_{X1}(.)$. Similarly, if the text modality is absent, $p_Y = 1.0$ and we get a sequence of \texttt{<mask>}s resulting in $\hat{E_A}$ being a sequence of mask embeddings.

Two adjustments must be made to allow for the addition of modalities: first, the modalities must have the same sequence length; if a modality is absent, its masked version is upsampled to the length of the other modality. Second, in order to match embedding sizes, each modality is passed through a learnable linear transform and layer normalization~\cite{lei2016layer} to match embedding sizes. This normalization is omitted from the above equation for brevity.

\noindent \textbf{Task-specific heads}: The final component of our model converts $Z$ into a model prediction. For text prediction, $Z$ needs to be converted into a sequence of logits used to compute the CTC loss. For speech prediction tasks, $Z$ is used to predict the log-mel features which are used to compute the L1 loss against the ground-truth. Formally,
\begin{align*}
    \begin{split}
        O_X &= W_X\text{SpeechHead}(Z) + b_X \\
        O_Y &= W_Y\text{TextHead}(Z) + b_Y
    \end{split}
\end{align*}
Here, $O_X$ and $O_Y$ are speech and text predictions respectively. SpeechHead$(.)$ and TextHead$(.)$ are modality specific non-autoregressive decoders modeled as 2-layer conformers. $W_X$, $W_Y$, $b_X$ and $b_Y$ are weights and biases of the prediction layer.

\subsection{Tasks}
\label{subsec:tasks}
We use our proposed model to perform six different tasks which include both supervised tasks with paired data and self-supervised tasks with unpaired data. Thus, given task-specific data $(X_{task}, Y_{task})$, the general forward propagation through our model is given below,
\begin{align}
\label{eqn:1}
    \begin{split}
        \hat{E}_A &= \text{DurMod}(\text{Mask}_{Y}(Y_{task},p_Y)) \\
        \hat{X} &= \text{Mask}_{X1/2}(X_{task},p_X,M) \\
        Z &= \text{MMEncoder}(\hat{X} + \hat{E_A}) \\
        O_X &= W_X\text{SpeechHead}(Z) + b_X \\
        O_Y &= W_Y\text{TextHead}(Z) + b_Y
    \end{split}
\end{align}
By controlling the task-specific data, the masking schedule and which task-specific head to use, we can define different modality-to-modality conversion tasks as explained below.

\noindent \textbf{Speech-to-text (STT)}: The STT task uses paired data, $(X_{STT}, Y_{STT})$. The input is speech, $X_{STT}$ and the desired output is the corresponding transcript, $Y_{STT}$. Thus, the text modality on the input side is completely masked, i.e. $p_Y=1.0$ in Equation~\ref{eqn:1}. The speech modality is completely unmasked, i.e. $p_X=0.0$. Finally, only the $\text{TextHead}(.)$ is used to compute the prediction logits, $O_{STT}$ from which we compute the STT loss as, $L_{STT} = \text{CTC-loss}(O_{STT}, Y_{STT})$. 

\noindent \textbf{Text-to-speech (TTS)}: TTS also uses paired data, $(X_{TTS}, Y_{TTS})$. Here, the input and output is text, $Y_{TTS}$ and speech, $X_{TTS}$ respectively. This time, the speech modality on the input side is fully masked, i.e. it is a sequence of zero-vectors with $p_X=1.0$. The predicted log-mels, $O_{TTS}$, are computed using the $\text{SpeechHead}(.)$ and the TTS loss, $L_{TTS} = \text{L1-Loss}(O_{TTS}, X_{TTS}) + \alpha L_{\text{dur}}$. Note that $L_{TTS}$ also includes the duration prediction loss.

\noindent \textbf{Text-to-text (T2T)}: Our model also allows for self-supervised tasks on unpaired data, the first of which is text-to-text, which we model as a masked language modeling task. For this we use unpaired text data, $Y_{T2T}$. For this task, the speech input modality is fully masked. The text modality is partially masked with $p_Y=0.25$ in Equation~\ref{eqn:1}. The TextHead$(.)$ is then used to compute the logits $O_{T2T}$ which is used to compute $L_{T2T} = \text{CTC-Loss}(\text{softmax}(O_{T2T}), Y_{T2T})$.

\noindent \textbf{Speech-to-Speech (S2S)}: The S2S task is similar to T2T. The input and output modalities are speech from unpaired speech data, $X_{S2S}$. The masking schedule used for speech is $\text{Mask}_{X1}(X_{S2S},p_X=0.0625,M=10)$ and the text is fully masked. $L_{S2S} = \text{L1-Loss}(O_{S2S}, X_{S2S})$ with $O_{S2S}$ is computed using SpeechHead$(.)$.

The next two tasks are useful for the iterative refinement procedure that we propose and explain in the next subsection.

\noindent \textbf{SpeechText-to-Text (ST2T)}: This task uses paired data $(X_{ST2T}, Y_{ST2T})$. For this task, the input is a combination of unmasked speech and partially masked text. Formally, $p_X=0.0$ and $p_Y \sim \{0.1, 0.25, 0.5, 0.75, 0.9\}$. The task is to predict the transcript, $Y_{ST2T}$ and the loss is computed as, $L_{ST2T} = \text{CTC-loss}(O_{ST2T}, Y_{ST2T})$.

\noindent \textbf{SpeechText-to-Speech (ST2S)}: This is similar to ST2T, using paired data $(X_{ST2S}, Y_{ST2S})$, except this time, the input text-modality is unmasked with $p_Y=0.0$. The schedule used to mask speech is $\text{Mask}_{X2}(X_{ST2S},p_X)$, where $p_X \sim \{0.1, 0.25, 0.5, 0.75, 0.9\}$. The loss, $L_{ST2S} = \text{L1-loss}(O_{ST2S}, X_{ST2S}) + \alpha L_{\text{dur}}$.

We train the model jointly using all six tasks together. The final loss is simply, $L_{\text{final}} = L_{STT} + L_{TTS} + L_{T2T} + L_{S2S} + L_{ST2T} + L_{ST2S}$.

\begin{figure}
    \hfill
    \centering
    \centerline{\includegraphics[width=\columnwidth]{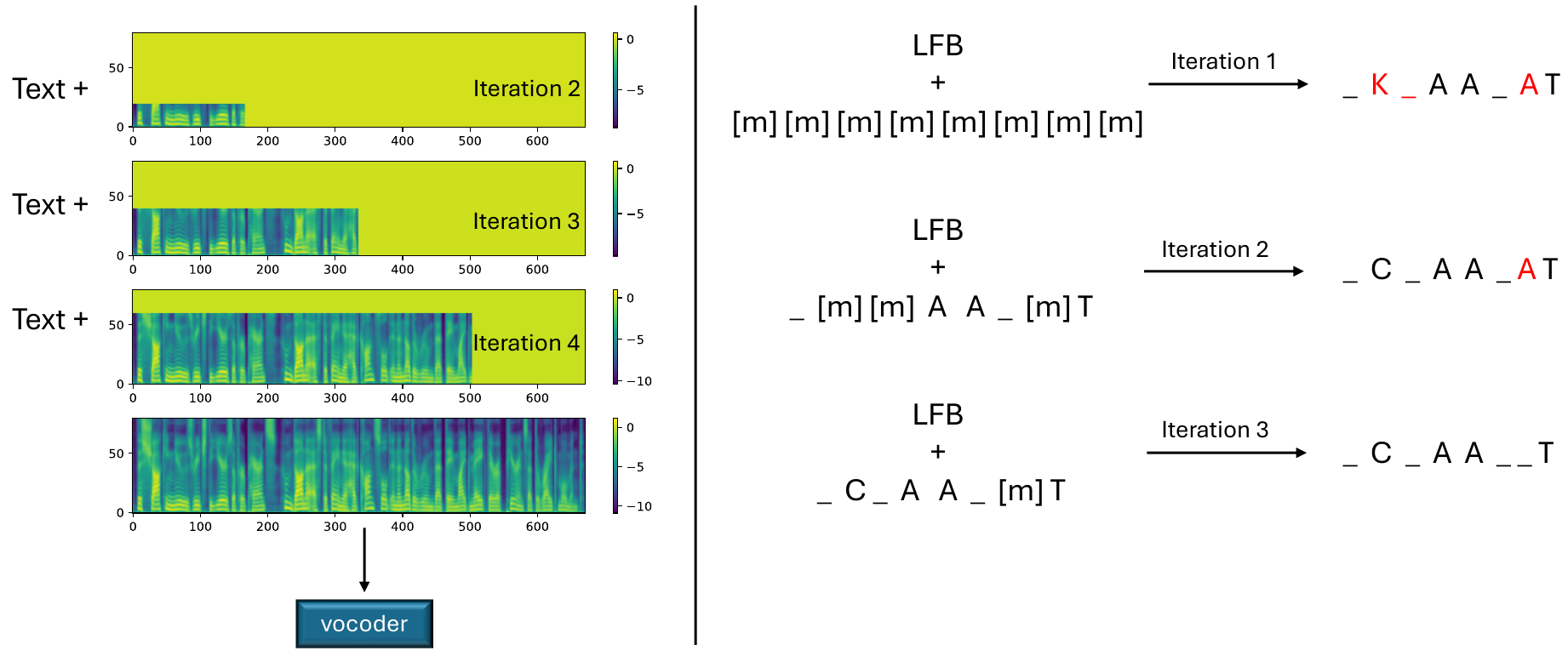}}
\caption{Iterative refinement illustration. For TTS (left), we show 4 iterations where the log-mel are gradually unmasked. For STT (right), the word \quotes{CAT} is to be predicted which is refined through 3 iterations. \quotes{[m]} refers to \texttt{<mask>}} 
\label{fig:iterative_ref}
\end{figure}

\subsection{Iterative refinement (Figure~\ref{fig:iterative_ref})}
\label{subsec:iterative refinement}

Our model is capable of performing multi-pass refinement of the desired output modality by masking the output during inference using some criterion and then feeding the masked output back to the input. This is where the ST2T and ST2S training criteria are useful: they involve a prediction that is conditioned on a masked version of itself.

For the STT task, we obtain a first-pass decoding by a forward pass. We obtain the confidence for each predicted character (excluding blanks) in the greedy CTC prediction by taking the average confidence of repetitions between consecutive blanks. Then, we mask the characters, their repetitions and the following blanks based on a confidence threshold. The partially masked hypothesis is fed back into the model for the next iteration of decoding where we follow the same steps. We perform $K$ iterations of this and the confidence threshold is linearly decayed from $0.99$ to $0.9$ over the $K$ steps. 


TTS can also benefit from a similar procedure; however, unlike text, log-mel prediction does not have a confidence measure. Thus, we gradually fill in the values along time and frequency over $K$ steps. In the $k^{th}$ iteration we unmask the first $k/K$ fraction of $T$ and $F$. 


\section{Experiments}
\label{sec:experiments}
\begin{table*}
     \centering
      \caption{\MakeLowercase{We use word error rate (WER) and UTMOS score~\cite{saeki2022utmos} for STT and TTS respectively. For TTS, we also report intelligibility and speaker similarity. Intelligibility of an audio is the WER of its transcription from the Whisper ASR model~\cite{radford2023robust}. This was 1.6\% and 2.0\% on the ground truth LJSpeech and LibriTTS. Speaker similarity is the cosine distance between speaker embeddings using~\cite{jung2024espnet}. } }
     \label{tab:main_results}
     \resizebox{\textwidth}{!}{
     \begin{tabular}{@{}lcccccccccccc@{}}\toprule
     && \multicolumn{3}{c}{LJSpeech (24 hours paired data)} && \multicolumn{7}{c}{LibriTTS (50 hours paired data)} \\
     \midrule
     Model && WER ($\downarrow$) & UTMOS ($\uparrow$) & Intel. ($\downarrow$) && WER ($\downarrow$) (dc) & WER ($\downarrow$) (tc) & WER ($\downarrow$) (do) & WER ($\downarrow$) (to) & UTMOS ($\uparrow$) & Intel. ($\downarrow$) & Spkr sim. ($\uparrow$) \\
     \midrule
     (1) STT only && 5.82 & - & - && 14.54 & 15.58 & 26.60 & 28.87 & - & - & - \\
     (2) TTS only && - & 3.82 & \textbf{1.7} && - & - & - & - & 3.72 & \textbf{3.10} & 0.754 \\ 
     (3) STT+TTS && 5.37 & 3.32 & 2.0 && 14.24 & 15.56 & 26.54 & 28.98 & 3.25 & 5.90 & 0.739 \\
     (4) STT+TTS+T2T+S2S && 5.33 & 3.47 & 2.0 && 13.54 & 14.90 & 25.48 & 27.93 & 3.31 & 4.30 & 0.741 \\
     (5) all six tasks && 5.07 & 3.69 & 1.9 && 12.73 & 14.06 & 24.49 & 26.97 & 3.51 & 3.80 & 0.754 \\
     (6) +Iterative refinement && \textbf{4.85} & \textbf{4.08} & 2.7 && \textbf{12.08} & \textbf{13.46} & \textbf{23.34} & \textbf{25.69} & \textbf{3.82} & 4.10 & \textbf{0.761} \\
     \bottomrule
     \end{tabular}
     }
 \end{table*}

 \begin{table}
     \centering
     \caption{\MakeLowercase{Results on Librispeech 100 hours. The intelligibility WER on the ground truth audio was 2.3\%}}
     \label{tab:results_librispeech}
     \resizebox{\columnwidth}{!}{
     \begin{tabular}{@{}lccccccc@{}}\toprule
          Model & WER (dc) & WER (tc) & WER (do) & WER (to) & UTMOS & Intel. &  Spkr sim. \\
          \midrule
          STT only & 10.17 & 10.75 & 27.68 & 28.23 & - & - & - \\
          TTS only & - & - & - & - & 3.66 & 3.7 & \textbf{0.506} \\
          all six tasks & 9.18 & 9.55 & 26.92 & 27.72 & 3.37 & \textbf{2.5} & 0.460 \\
          +Iterative refinement & \textbf{8.56} & \textbf{8.78} & \textbf{25.15} & \textbf{25.75} & \textbf{3.68} & 2.6 & 0.466 \\
          \bottomrule
     \end{tabular}}
 \end{table}

 \begin{table}
     \centering
     \caption{\MakeLowercase{Mean opinion score (MOS) on a 40 sample set of LibriTTS test-clean. All models trained on 50 hours of paired speech.}}
     \label{tab:human_eval}
     \resizebox{0.5\columnwidth}{!}{
     \begin{tabular}{@{}lc@{}}\toprule
         Model & MOS ($\uparrow$) \\
         \midrule
         Ground truth & 4.37 $\pm$ 0.05 \\
         \midrule
         TTS only & 2.96 $\pm$ 0.07 \\
         STT+TTS & 2.26 $\pm$ 0.06 \\
         all six tasks & 2.43 $\pm$ 0.06 \\
         +Iterative refinement & 2.84 $\pm$ 0.07 \\
         \bottomrule
     \end{tabular}}
 \end{table}

\begin{figure}
    \hfill
    \centering
    \centerline{\includegraphics[width=\columnwidth]{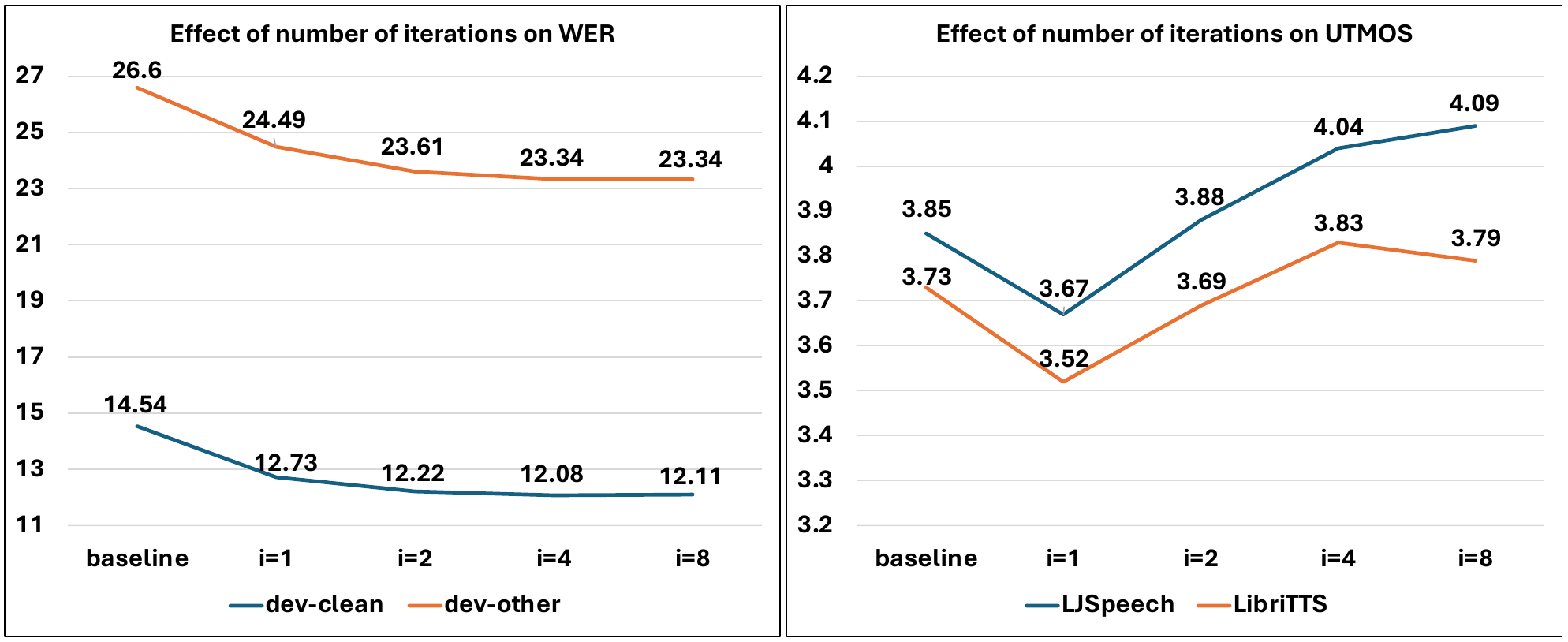}}
\caption{Effect of number of iterations of iterative refinement on STT (left) and TTS (right). These results are on the development sets.} 
\label{fig:chart_iter}
\end{figure}
\subsection{Data}
\label{subsec:data}
\subsubsection{LJSpeech~\cite{ito2017lj}}
This is a single speaker dataset with audio amounting to roughly 24 hours. It contains 13,100 audio clips from which we randomly take 12,500 samples for training, 300 samples for validation and 300 samples for testing. Note that for this data, we use the same set of 12,500 samples for the supervised and self-supervised tasks, i.e. we do not have a separate set of unpaired data.

\subsubsection{LibriTTS~\cite{zen2019libritts}}
This dataset contains speech from multiple speakers (around 2000) and is a subset of Librispeech~\cite{panayotov2015librispeech}. The total data amounts to around 550 hours. We use the 50 hour \quotes{clean-50} subset as the paired data, 200 hour \quotes{clean-200} as the unpaired speech data and 300 hour \quotes{other-300} as the unpaired text data. We used the restored version of this dataset~\cite{koizumi2023libritts}. STT results are reported on the dev-clean (dc), test-clean (tc), dev-other (do) and test-other (to) subsets and TTS results are reported on the test-clean subset.

\subsubsection{Librispeech~\cite{panayotov2015librispeech}}
We also use the original Librispeech corpus. The train-clean-100 subset served as the paired data, train-clean-360 as the unpaired speech and train-other-500 as the unpaired text. 

\subsection{Training details}
 All modules in the model are conformer blocks~\cite{gulati2020conformer} with either 256 hidden units and 4 attention heads (for LJSpeech) or 384 hidden units and 6 attention heads (for LibriTTS and Librispeech). The MMEncoder has 12 layers and the rest have 2 layers each.

 For multi-speaker TTS, we make use of pretrained speaker embeddings from the ECAPA-TDNN model~\cite{desplanques2020ecapa}. These are added to the alignment embeddings, $E_A$ in the duration model for multispeaker data. For TTS, the model predicts 80-dimensional LFB features which are converted to the speech waveform using the HiFi-GAN vocoder~\cite{kong2020hifi}.

 When using text-only data, we do not have true CTC alignments to use for the T2T task. We train a standalone duration model using the paired data which is used to generate pseudo-CTC alignments.

 For STT, we use characters at the output vocabulary and 80-dimensional LFB features as input. Only when using $(X_{STT}, Y_{STT})$, we apply data augmentation in the form of SpecAugment~\cite{park2019specaugment} and speed and tempo augmentation~\cite{ko2015audio} with factors of 0.9 and 1.1. The augmented data is only used for the STT task.

 All joint models are trained for 100~epochs with a OneCycleLR annealing policy~\cite{smith2019super} of the learning rate whose peak value is 5e-4. A linear annealing happens after 30 epochs of warmup. We scale the peak learning rate with batch size as $lr = 0.0005\times\sqrt{\frac{bsize}{32}}$ 

 \subsection{Results and discussion}
 
\noindent \textbf{Main results}: Table~\ref{tab:main_results} shows the results on LJSpeech and LibriTTS, two commonly used datasets for TTS. Rows (1) and (2) are the task-specific baselines. Rows (3) to (6) show the results of the proposed models as we perform STT and TTS jointly (row (3)), add self-supervised tasks (row (4)) and add multimodal input based supervised tasks (ST2T and ST2S) with iterative refinement (rows (5) and (6)). 

When training STT and TTS jointly without additional tasks (row~(3)), we see that while the STT performance does not get affected as much, the TTS performance is deteriorated compared to the TTS-only baseline. We hypothesize that STT and TTS are not complementary tasks with the former being many-to-one and the latter being one-to-many. While STT may be able to recover its performance in the presence of a non-complementary task, it may be much harder for TTS to do so. With the addition of self-supervised tasks like T2T and S2S, we see that TTS performance starts to improve (row (4)) as evidenced by the UTMOS score. In row (5), we show that the ST2T and ST2S tasks improve performance across the board and with the use of iterative refinement (row (6)), we obtain the best performance of our joint models. We note that ST2T and ST2S are useful multimodal tasks which perhaps help with further data augmentation.

With iterative refinement, we are able to achieve the best STT performance on both datasets. For TTS, we are able to outperform the baseline in terms of UTMOS score. In terms of intelligibility, iterative refinement does not give an improvement but the addition of ST2T and ST2S tasks helps us to get close to the baseline performance. In terms of speaker similarity on LibriTTS, the proposed model with iterative refinement is able to outperform the baseline performance. 

Next, we see how our model performs on the Librispeech dataset which is primarily used for STT in literature. However, we evaluate our models on Librispeech for both STT and TTS tasks. We use the 100-hour clean subset as the paired data. The results in Table~\ref{tab:results_librispeech} show that similar to Table~\ref{tab:main_results}, the STT results are improved with the addition of tasks and the iterative refinement strategy, with the TTS performance being very close to the baseline performance. Librispeech is a challenging dataset from a TTS perspective, hence we see an overall subpar performance. However, note that all of our proposed models can perform both STT and TTS tasks with better or similar performance compared to the baseline task specific models. 

\noindent \textbf{Effect of iterations}: We show the effect of number of iterations, $K$, on the STT and TTS performance using the iterative refinement approach in Figure~\ref{fig:chart_iter}. Note that the performance on both tasks keeps improving until $K=4$ after which the performance saturates.

\noindent \textbf{Human evaluation}: We also show human evaluations on the synthesized speech of all our models in Table~\ref{tab:human_eval} on a 40-sample set of the LibriTTS test-clean data. Compared to the UTMOS scores in Table~\ref{tab:main_results}, the MOS scores are low, indicating that the speech produced is not very high quality. We emphasize that although this is the case, the evaluated TTS model was trained with only 50 hours of paired data for a difficult multi-speaker TTS task. Also note that the addition of tasks defined in Section~\ref{subsec:tasks} improves the quality of the audio to some extent in our proposed joint model. All our model development was done using UTMOS score as a guide since it is an automated metric, much easier to compute compared to the MOS evaluation.

\section{Limitations and conclusion}
Although our proposed model can perform both STT and TTS, we believe that the performance on these tasks can be improved further. For STT, we can use techniques like label smoothing~\cite{szegedy2016rethinking}, DropConnect~\cite{wan2013regularization}, sequence noise injection~\cite{saon2019sequence} and self-conditioned CTC~\cite{nozaki2021relaxing}. For TTS, we may use FastSpeech2's~\cite{ren2020fastspeech} approach of predicting pitch and energy as intermediate features. We can also utilize the entire 550 hours of paired data from LibriTTS instead of just the 50 hours. In addition, using a phoneme-based alignment as input instead of a character-level alignment can also help TTS.


Our proposed model does open numerous possibilities for modeling various tasks like STT, TTS, S2ST, S2TT and SLU in a non-autoregressive manner. The fact that STT and TTS can indeed be modeled together as shown by our work in addition to the proposed improvements using unpaired data and iterative refinement is an encouraging step in building larger multimodal models.

\section*{Acknowledgement}
We thank Avihu Dekel and Nimrod Shabtay of IBM Research, Israel, for their work on the speaker similarity measurement tool. We also thank Ankit Gupta from IBM Research, NY, USA, for his work on data creation.

\bibliographystyle{IEEEtran}
\bibliography{IEEEexample}

\end{document}